# Intelligent Financial Fraud Detection Practices: An Investigation

Jarrod West[1], Maumita Bhattacharya and Rafiqul Islam

School of Computing & Mathematics
Charles Sturt University, Australia
`jnwest@netspace.net.au`[1], `{mbhattacharya,mislam}@csu.edu.au`

**Abstract.** Financial fraud is an issue with far reaching consequences in the finance industry, government, corporate sectors, and for ordinary consumers. Increasing dependence on new technologies such as cloud and mobile computing in recent years has compounded the problem. Traditional methods of detection involve extensive use of auditing, where a trained individual manually observes reports or transactions in an attempt to discover fraudulent behaviour. This method is not only time consuming, expensive and inaccurate, but in the age of big data it is also impractical. Not surprisingly, financial institutions have turned to automated processes using statistical and computational methods. This paper presents a comprehensive investigation on financial fraud detection practices using such data mining methods, with a particular focus on computational intelligence-based techniques. Classification of the practices based on key aspects such as detection algorithm used, fraud type investigated, and success rate have been covered. Issues and challenges associated with the current practices and potential future direction of research have also been identified.

## 1 Introduction and Background

Financial fraud is an issue that has wide reaching consequences in both the finance industry and daily life. Fraud can reduce confidence in industry, destabilise economies, and affect people's cost of living. Traditional approaches of fraud detection relied on manual techniques such as auditing, which are inefficient and unreliable due to the complexities associated with the problem. Computational intelligence (CI)-based as well as conventional data mining approaches have been proven to be useful because of their ability to detect small anomalies in large data sets [14].

Financial fraud is a broad term with various potential meanings, but for our purposes it can be defined as the intentional use of illegal methods or practices for the purpose of obtaining financial gain [30]. There are many different types of financial fraud, as well as a variety of data mining methods, and research is continually being undertaken to find the best approach for each case. The common financial fraud cate-gories and the popular data mining as well as computational intelligence-based techniques used for financial fraud detection are depicted in Fig. 1 and Fig. 2 respectively.

Advancements in modern technologies such as the internet and mobile computing have led to an increase in financial fraud in recent years [27]. Social factors such as the increased distribution of credit cards have increased spending, but also resulted in an increase to fraud [20]. Fraudsters are continually refining their methods, and as such there is a requirement for detection methods to be able to evolve accordingly. CI and data mining have already been shown to be useful in similar domains such as credit card approval, bankruptcy prediction, and analysis of share markets [16]. Fraud detection is primarily considered to be a classification problem, but with a vast imbalance in fraudulent to legitimate transactions misclassification is common and can be significantly costly [6]. Many data mining approaches are efficient classifiers and are applicable to fraud detection for their efficiency at processing large datasets and their ability to work without extensive problem specific knowledge [19].

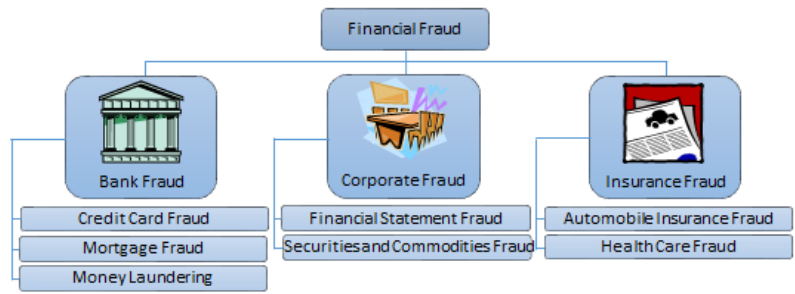

**Fig. 1.** Common financial fraud categories.

A useful framework for applying CI or data mining to fraud detection is to use them as methods for classifying suspicious transactions or samples for further consideration. Studies show that reviewing 2% of credit card transactions could reduce fraud losses to 1% of the total cost of all purchases, with more assessments resulting in smaller loss but with an increase in auditing costs [18]. A multi-layer pipeline approach can be used with each step applying a more rigorous method to detect fraud. Data mining can be utilised to efficiently filter out more obvious fraud cases in the initial levels and leave the more subtle ones to be reviewed manually [18].

Early fraud detection studies focused on statistical models such as logistic regression, as well as neural networks (see [18], [28] and [9] for details). In 1995 Sohl et al. first predicted financial statement fraud using a back-propagation neural network [28]. More recently, in addition to examining financial scenarios such as stock market and bankruptcy prediction, Zhang et al. applied various data mining techniques to financial fraud detection in 2004 [29]. In 2005 Vatsa et al. investigated a novel approach using game theory which modelled fraudsters and detection methods as opposing players in a game, each striving to obtain the greatest financial advantage [22]. A process mining approach was used by Yang et al. in 2006 to detect health care fraud [26]. In 2007 Yue et al. observed that, to date, classification-based methods are both the most commonly researched techniques as well as the only successful ones [28]. The chronological progression of some of the recent financial fraud detection research has been depicted in Fig. 3.

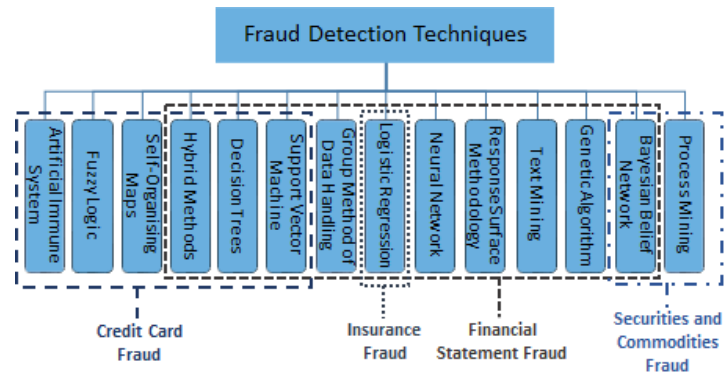

**Fig. 2.** Detection algorithms used for various fraud categories.

In this paper we provide a comprehensive investigation of the existing practices in financial fraud detection. We present a detailed classification of such practices; aimed at informing development of enhanced financial fraud detection frameworks. The remainder of the paper is structured as follows: Section 2 presents a comprehensive classification of the existing practices in financial fraud detection based on fraud type, detection algorithm, success rate and so on. Section 3 offers an insight into issues and challenges associated with financial fraud detection and potential direction for future research. Finally, Section 4 presents some concluding remarks.

## 2 Classification of Financial Fraud Detection Practices

In the following sub-sections we will classify existing financial fraud detection practices based on success rate, detection technique used, and fraud type. This categorisation will enable us to identify trends in current practices, including which have been successful, probable factors influencing the outcomes, and also any gaps in the research.

### 2.1 Classification Based on Performance

A variety of standards have been used to determine performance, but the three most commonly used are *accuracy*, *sensitivity*, and *specificity*. Accuracy measures the ratio of all successfully classified samples to unsuccessful ones. Sensitivity compares the amount of items correctly identified as fraud to the amount incorrectly listed as fraud, also known as the ratio of true positives to false positives. Specificity refers to the same concept with legitimate transactions, or the comparison of true negatives to false negatives [3], [19].

Tables 1, 2, and 3 classify financial fraud detection research based on these performance measures. Additionally, Fig 4 depicts the broad comparative performance of various fraud detection methods.

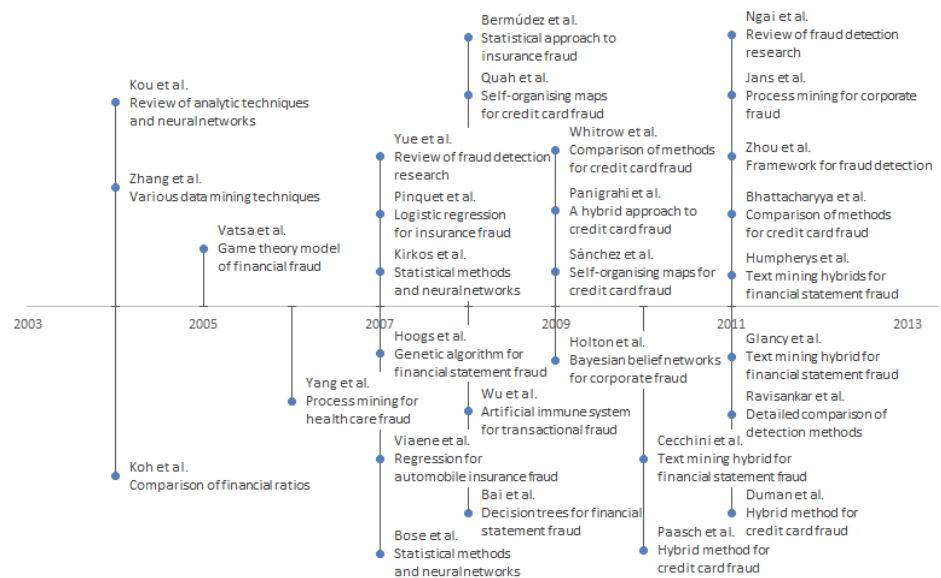

**Fig. 3.** Chronological progression of recent financial fraud detection research.

In addition to the three performance measures discussed here, several other performance measures have been used in the literature. For example, Duman et al. chose to show their results for sensitivity in graph form instead of deterministic values, grouped by each set of input parameters [6]. In addition to other forms of graphing [18], some research used software-determined success levels or case-based procedures to determine the success of their fraud detection techniques [20], [11].

From the results we can see that CI methods typically had better success rate than statistical methods. Sensitivity was slightly better for random forests and support vector machines than logistic regression, with comparable specificity and accuracy [3]. Genetic programming, support vector machines, probabilistic neural networks, and group method of data handling outperformed regression in all three areas [19]. Additionally, a neural network with exhaustive pruning was found to be more specific and accurate than CDA [4]. One statistical method seems to contradict this theory however: Bayesian belief networks were reported to be more accurate than neural networks and decision trees [12].

Most of the research showed a large difference between each method's sensitivity and specificity results. For example, Bhattacharyya et al. showed that logistic regression, support vector machines and random forests all performed significantly better at detecting legitimate transactions correctly than fraudulent ones [3]. Support vector machines, genetic programming, neural networks, group method of data handling, and particularly logistic regression were also slightly less sensitive [19]. Also a neural network with exhaustive pruning showed more specificity than sensitivity [4].

As explained previously, fraud detection is a problem with a large difference in misclassification costs: it is typically far more expensive to misdiagnose a fraudulent transaction as legitimate than the reverse. With that in mind it would be beneficial for detection techniques to show a much higher sensitivity than specificity, meaning that these results are less than ideal. Contrary to this belief, Hoogs et al. hypothesised that financial statement fraud may carry higher costs for false positives, and their results reflect this with a much higher specificity [9]. Panigrahi et al. also acknowledged the costs associated with following up credit card transactions marked as fraudulent, focussing their results on sensitivity only [16]. The CDA and CART methods, as well as neural networks, Bayesian belief networks and decision trees performed better in this regard, with all showing a somewhat higher ability to classify fraudulent transactions than legitimate ones [4], [12].

**Table 1.** Accuracy results for fraud detection practices

| Research | Fraud Investigated | Method Investigated | Accuracy |
|---|---|---|---|
| [3] | Credit card transaction fraud from a real world example | Logistic model (regression) | 96.6-99.4% |
| | | Support vector machines | 95.5-99.6% |
| | | Random forests | 97.8-99.6% |
| [12] | Financial statement fraud from a selection of Greek manufacturing firms | Decision trees | 73.6% |
| | | Neural networks | 80% |
| | | Bayesian belief networks | 90.3% |
| [19] | Financial statement fraud with financial items from a selection of public Chinese companies | Support vector machine | 70.41-73.41% |
| | | Genetic programming | 89.27-94.14% |
| | | Neural network (feed forward) | 75.32-78.77% |
| | | Group method of data handling | 88.14-93.00% |
| | | Logistic model (regression) | 66.86-70.86% |
| | | Neural network (probabilistic) | 95.64-98.09% |
| [7] | Financial statement fraud with managerial statements for US companies | Text mining with singular validation decomposition vector | 95.65% |
| [5] | Financial statement fraud with managerial statements for US companies | Text mining | 45.08-75.41% |
| | | Text mining and support vector machine hybrid | 50.00-81.97% |
| [10] | Financial statement fraud with managerial statements for US companies | Text mining and decision tree hybrid | 67.3% |
| | | Text mining and Bayesian belief network hybrid | 67.3% |
| | | Text mining and support vector machine hybrid | 65.8% |
| [4] | Financial statement fraud with financial items from a selection of public Chinese companies | CDA | 71.37% |
| | | CART | 72.38% |
| | | Neural network (exhaustive pruning) | 77.14% |

*Remarks:* Considering the three performance measures, namely accuracy, sensitivity and specificity, our investigation shows that the computational intelligence-based approaches have generally performed better than the statistical approaches in most cases.

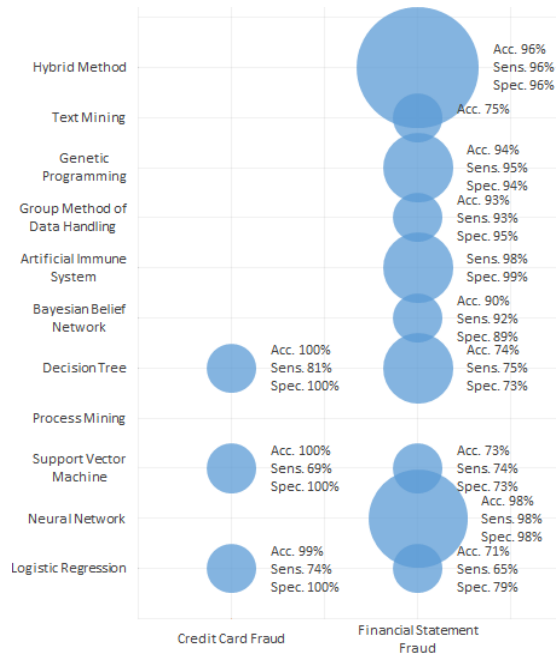

**Fig. 4.** Comparative performance of various detection methods.

### 2.2 Classification Based on Detection Algorithm

Classifying fraud detection practices by the detection algorithm used is a useful way to identify the suitable techniques for this problem domain. It can also help us to determine why particular methods were chosen or successful. Additionally, we can identify any gaps in research by looking at algorithms which have not been explored sufficiently. Table 4 shows classification of financial fraud detection practices based on detection algorithm (conventional data mining and CI-based approaches) used.

Previously it was mentioned that early fraud detection research focussed on statistical models and neural networks; however, it may be noted that these methods still continue to be popular. Many used at least one form of neural network [12], [19], [4], some investigated logistic regression [3], [17], [23], [19], while others applied Bayesian belief networks [8], [12], [2]. Application of CDA has been relatively uncommon [4]. Neural networks and logistic regression are often chosen for their well-established popularity, giving them the ability to be used as a control method by which other techniques are tested. Comparatively, more advanced methods such as support vector machines and genetic programming have received substantially less attention. Yue et al. also reported that all the methods mentioned in their research were a form of classification, with no studies performed on clustering or time-series approaches, and that most of the research focussed on supervised learning as opposed to unsupervised [28].

**Table 2.** Sensitivity results for fraud detection practices

| Research | Fraud Investigated | Method Investigated | Sensitivity |
|---|---|---|---|
| [3] | Credit card transaction fraud from a real world example | Logistic model (regression) Support vector machines Random forests | 24.6-74.0% 43.0-68.7% 42.3-81.2% |
| [12] | Financial statement fraud from a selection of Greek manufacturing firms | Decision trees Neural networks Bayesian belief networks | 75.0% 82.5% 91.7% |
| [19] | Financial statement fraud with financial items from a selection of public Chinese companies | Support vector machine Genetic programming Neural network (feed forward) Group method of data handling Logistic model (regression) Neural network (probabilistic) | 55.43-73.60% 85.64-95.09% 67.24-80.21% 87.44-93.46% 62.91-65.23% 87.53-98.09% |
| [7] | Financial statement fraud with managerial statements | Text mining with singular validation decomposition vector | 95.65% |
| [4] | Financial statement fraud with financial items from a selection of public Chinese companies | CDA CART Neural network (exhaustive pruning) | 61.96% 72.40% 80.83% |
| [16] | Credit card fraud using legitimate customer transaction history as well as generic fraud transactions | Bayesian learning with Dempster-Shafer combination | 71-83% |
| [9] | Financial statement fraud from Accounting and Auditing Enforcement Releases by the Securities and Exchange Commission | Genetic algorithm | 13-27% |
| [25] | Transactional fraud in automated bank machines and point of sale from a financial institution | Coevolution artificial immune system Standard evolution artificial immune system | 97.688-98.266% 92.486-95.376% |

Several of the research focussed on a single form of fraud detection which they advocated above others, such as studying text mining with the singular validation decomposition vector [7], self-organising maps [18], logistic regression [23], [17], and fuzzy logic [20]. Additionally, some researchers focussed soley on classification and regression trees [1], Bayesian belief networks [8], individual statistical techniques [16], or their own hybrid methods [6]. This unilateral approach is useful for demonstrating the ability of the specific method in isolation, but without comparing it to other methods it is difficult to understand the relative performance of the technique. Additional factors such as the fraud type researched and the specific dataset used can influence the results of the experiment. Future research could focus on reviewing these methods against other more established techniques.

**Table 3.** Specificity results for fraud detection practices

| Research | Fraud Investigated | Method Investigated | Specificity |
| --- | --- | --- | --- |
| [3] | Credit card transaction fraud from a real world example | Logistic model (regression) | 96.7-99.8% |
| | | Support vector machines | 95.7-99.8% |
| | | Random forests | 97.9-99.8% |
| [12] | Financial statement fraud from a selection of Greek manufacturing firms | Decision trees | 72.5% |
| | | Neural networks | 77.5% |
| | | Bayesian belief networks | 88.9% |
| [19] | Financial statement fraud with financial items from a selection of public Chinese companies | Support vector machine | 70.41-73.41% |
| | | Genetic programming | 89.27-94.14% |
| | | Neural network (feed forward) | 75.32-78.77% |
| | | Group method of data handling | 88.34-95.18% |
| | | Logistic model (regression) | 70.66-78.88% |
| | | Neural network (probabilistic) | 94.07-98.09% |
| [7] | Financial statement fraud with managerial statements | Text mining with singular validation decomposition vector | 95.65% |
| [4] | Financial statement fraud with financial items from a selection of public Chinese companies | CDA | 80.77% |
| | | CART | 72.36% |
| | | Neural network (exhaustive pruning) | 73.45% |
| [9] | Financial statement fraud from Accounting and Auditing Enforcement Releases by the Securities and Exchange Commission | Genetic algorithm | 98%-100% |
| [25] | Transactional fraud in automated bank machines and point of sale from a financial institution | Coevolution artificial immune system | 95.862-97.122% |
| | | Standard evolution artificial immune system | 99.311% |

A rising trend in fraud detection is the use of hybrid methods which utilise the strengths of multiple algorithms to classify samples. Duman and Ozcelik used a combination of scatter search and genetic algorithm, based on the latter but targeting attributes of scatter search such as the smaller populations and recombination as the reproduction method [6]. A different approach was taken by Panigrahi et al. who used two methods sequentially, beginning with the Depster-Schaefer method to combine rules and then using a Bayesian learner to detect the existence of fraud [16]. Some researchers applied fuzzy logic to introduce variation to their samples, attempting to transform it to resemble real world data before deploying a different technique to actually detect the presence of fraud [11]. The investigators recognised that applying 'fuzziness' to their problem increased the performance of their solution [25]. Similarly, several researchers combined traditional computational intelligence methods with text mining to analyse financial statements for the presence of fraud [5], [10].

*Remarks:* Based on our investigation, it is apparent that neural networks and statistical algorithms have continued to remain popular through recent years, while hybrid methods are a rising trend in financial fraud detection, combining the strengths of multiple techniques.

## 2.3 Classification Based on Fraud Type

Given the varying nature of each type of fraud, the problem domain can differ significantly depending on the form that is being detected. By classifying the existing practices on the type of fraud investigated we can identify the techniques more suitable and more commonly used for a specific type of fraud. Additionally we can infer the varieties which are considered the most important for investigation depending on the scope and scale of their impact. Table 5 depicts the classification based on fraud types considered, along with the detection methods used.

With each chosen algorithm, feature selection will differ depending on the problem domain. Specific financial statement fraud exists within individual companies, and as such attribute ratios are used instead of absolute values. Koh and Low provide a good example of the relevant ratios such as net income to total assets, interest payments to earnings before interest and tax, and market value of equity to total assets [13]. In comparison, research into credit card fraud has typically selected independent variables or aggregate values which may be quantitative or qualitative. For example, Bhattacharyya et al. made use of transaction amount, categorical values such as account number, transaction date, and currency, and aggregated properties like total transaction amount per day, and average amount spent at a single merchant [3].

We can see that the existing research has been greatly unbalanced in fraud type studied. The vast majority of research has focussed on two forms of financial fraud: *credit card fraud* and *financial statement fraud*. Only a handful of studies have looked at securities and commodities fraud; also many studies focus on external forms of corporate fraud while neglecting the internal ones [11]. Ngai et al. found that insurance fraud had received the highest coverage during their research [14]: the fact that we identified only a few examples of published literature on this type of fraud since 2007 indicates that research into insurance fraud is declining. Additionally, no studies have been performed directly on mortgage fraud or money laundering. The reason for this disparity may be the differing relevance to stakeholders of each fraud type.

*Remarks:* Through our investigation we observe a significant imbalance in fraud type studied, with the majority focussing on either financial statement fraud or credit card fraud. Other forms of corporate fraud have received little attention, and hardly any studies have been done into mortgage fraud or money laundering.

Table 4. Classification based on detection algorithm used

| Method Investigated | Relevant Method Properties | Fraud Investigated | Research |
|---|---|---|---|
| Neural network | Capable of adapting to new trends, able to handle problems with no algorithmic solution. Typically used for classification and prediction. | Financial statement fraud | [4], [12], [19] |
| Logistic model | Suitable for categorical classification problems like fraud detection. Typically used for regression. | Credit card fraud  Insurance fraud  Financial statement fraud | [3]  [17], [23], [2]  [19] |
| Support vector machine | Able to handle unbalanced data and complicated relationships between variables. Typically used for classification and prediction. | Credit card fraud  Financial statement fraud | [3], [24]  [19] |
| Decision trees, forests and CART | Easy to use and has a well-documented ability with similar problems. Typically used for classification and prediction. | Credit card fraud  Financial statement fraud | [3], [24]  [4], [1], [12] |
| Genetic algorithm/programming | Suitable for binary classification as the fitness function can be the accuracy of the population. Typically used for classification. | Financial statement fraud | [19], [9] |
| Text mining | Capable of studying plain text, which offers a new dimension to the problem. Typically used for clustering and anomaly detection. | Financial statement fraud | [7], [5] |
| Group method of data handling | Provides many of the same benefits as neural networks. Typically used for prediction. | Financial statement fraud | [19] |
| Response-surface methodology | Useful for determining which method is best applied to the problem domain. | Financial statement fraud | [29] |
| Self-organizing map | Provide both clustering and classification abilities, similarly to neural networks. Typically used for classification and clustering. | Credit card fraud | [18], [20] |
| Bayesian belief net- | Structured and formulaic, used extensively in other problems | Insurance fraud  Corporate fraud | [2]  [8] |
| work | with good results. Typically used for prediction and anomaly detection. | Financial statement fraud | [12] |
| Process mining | Objective and able to work well with large samples of existing data. Typically used for anomaly detection. | Securities and commodities fraud | [11] |
| Artificial immune system | Utilises binary matching rules, shown to be very powerful when paired with fuzzy logic. Typically used for anomaly detection. | Credit card fraud | [25] |
| Hybrid methods | Combines the strengths of multiple standard algorithms into a new, superior method. Can be used for any combination of classification, clustering, prediction, regression, and anomaly detection. | Credit card fraud Financial statement fraud | [16], [6] [5], [10] |
| All/generic | Allows the comparison of multiple methods on a specific problem to discover the benefits and negatives of each. Can be used for any combination of classification, clustering, prediction, regression, and anomaly detection. | All/generic | [28], [14] |

## 3   Financial Fraud Detection: Challenges and Future Directions

Financial fraud detection is an evolving field in which it is desirable to stay ahead of the perpetrators. Additionally, it is evident that there are still facets of intelligent fraud detection that have not been investigated. In this section we present some of the key issues associated with financial fraud detection and suggest areas for future research. Some of the identified issues and challenges are as follows:

- *Typical classification problems:* CI and data mining-based financial fraud detection is subject to the same issues as other classification problems, such as feature selection, parameter tuning, and analysis of the problem domain.
- *Fraud types and detection methods:* Financial fraud is a diverse field and there has been a large imbalance in both fraud types and detection methods studied: some have been studied extensively while others, such as hybrid methods, have only been looked at superficially.
- *Privacy considerations:* Financial fraud is a sensitive topic and stakeholders are reluctant to share information on the subject. This has led to experimental issues such as undersampling.
- *Computational performance:* As a high-cost problem it is desirable for financial fraud to be detected immediately. Very little research has been conducted on the computational performance of fraud detection methods for use in real-time situations.
- *Evolving problem:* Fraudsters are continually modifying their techniques to remain undetected. As such detection methods are required to be able to constantly adapt to new fraud techniques.
- *Disproportionate misclassification costs:* Fraud detection is primarily a classification problem with a vast difference in misclassification costs. Research on the performance of detection methods with respect to this factor is an area which needs further attention.
- *Generic framework:* Given that there are many varieties of fraud, a generic framework which can be applied to multiple fraud categories would be valuable.

**Table 5.** Classification based on fraud type investigated

| Fraud Type | Method Applied | Research on the Type of Fraud |
|---|---|---|
| Credit card | Support vector machines; Decision tree; Self-organising maps; Fuzzy logic; Artificial immune system; Hybrid methods | [3] investigated credit card fraud from an international operation; [18] investigated a banking database from the Singapore branch of a well-known international bank; [20] investigated fraud in multinational department stores; [6] investigated typical consumer spending to determine fraud in a major bank in Turkey; [16] investigated variation in legitimate customer transaction behaviour with synthesised credit card data; [25] investigated automated bank machines and point of sale from an anonymous financial institution; [24] investigated credit card transactions. |
| Securities and commodities and other Corporate | Bayesian belief network; Process mining | [11] investigated internal transactional fraud from a successful, anonymous European financial institution; [8] Investigated emails and discussion group messages to detect corporate fraud. |
| Insurance Fraud | Logistic model | [17], [23] and [2] all investigated motor insurance claims from Spanish insurance companies. |
| Financial statement | Response-surface methodology; Neural networks; Decision trees; Bayesian belief networks; Support vector machine; Genetic algorithms; Group method of data handling; Logistic model (regression); Text mining; Hybrid methods | [29] investigated financial statement fraud in general; [12] investigated a selection of Greek manufacturing firms; [19], [4], and [1] investigated a series of public Chinese companies; [7] and [10] investigated managerial statements from official company documents; [9] and [5] investigated Accounting and Auditing Enforcement Releases authored by a selection of US companies. |

As a classification problem, financial fraud detection suffers from the same issues as other similar problems. Feature selection has a high impact on the success of any classification method. While some researchers have mentioned feature selection for one type of fraud [13], [3], no comparisons have been made between features for differing problem domains. Also, one of the major benefits of the computational intelligence and data mining methods is their ability to be adjusted to fit the problem domain. Existing research has rarely used any form of customisation or tuning for specific problems; however, tuning is an important factor in the context of an algorithm's performance. For example, the number of nodes and internal layers within a neural network has a large impact on both accuracy and computational performance. Similarly the kernel function chosen will considerably alter the success of a support vector machine and parameters such as the fitness function, crossover method, and probability for mutation will impact the results of a genetic programming algorithm. Research on customisation or tuning of the computational methods is required to truly comprehend the ability of each method. Further, in other data mining cases the solution algorithm is selected based on its performance within the problem domain, which for financial fraud detection is the type of fraud investigated. Studies on the suitability of various methods for each fraud category are necessary to understand which attributes of each algorithm make them appropriate for detecting financial fraud.

From the existing literature it is apparent that there are some forms of fraud that have not been investigated as extensively as others. Financial statement fraud has been considerably investigated, which is understandable given its high profile nature, but there are other forms of fraud that have a significant impact on consumers. Credit card fraud often has a direct impact on the public and the recent increase in online transactions has led to a majority of the U.S. public being concerned with identity theft [3]. A benefit of this close relation to the user is that credit card fraud is typically detected quickly, which gives researchers access to large datasets of unambiguous transactions. Other forms of fraud which have not been covered in depth include money laundering, mortgage, and securities and commodities fraud. A lack of sufficient sample size may be the reason for the lack of research in these areas [14]. Future studies that focussed on these types of fraud detection would be beneficial.

The private nature of financial data has led to institutions being reluctant to share fraudulent information. This has had an affect both on the fraud types that have been investigated as well as the datasets used for the purpose. In the published literature many of the financial fraud simulations consisted of less than a few hundred samples, typically with comparable amounts of fraudulent and legitimate specimens. This is contrary to the realities of the problem domain, where fraud cases are far outweighed by legitimate transactions [3]. Undersampling the problem domain like this can cause biases in the data that do not accurately represent real-world scenarios [9]. There is a definite need for further studies with realistic samples to accurately depict the performance of each method [7].

Some forms of financial fraud occur very rapidly, such as credit card fraud. If a fraudster obtains an individual's credit card information it's very likely that they will use it immediately until the card limit is reached. The ability to detect fraud in real-time would be highly beneficial as it may be able to prevent the fraudster from making subsequent transactions. Computational performance is therefore a key factor to consider in fraud detection. Though some researchers have noted the performance of their particular methods [3], [18], most studies were simulations performed on test datasets. Further research focussing on the computational as well as classification performance is required.

Unlike many classification problems, fraud detection solutions must be capable of handling active attempts to circumvent them. As detection methods become more intelligent, fraudsters are also constantly upgrading their techniques. For example, in the last few decades credit card fraud has moved from individuals stealing or forging single cards to large-scale phone and online fraud perpetrated by organised groups [3]. It is therefore necessary for fraud detection methods to be capable of evolving to stay ahead of fraudsters. Some researchers have considered models for adaptive classification, however further research is required to fully develop these for use in practical fraud detection problems [30].

As explained previously fraud has a large cost to businesses. Additionally, fraud detection has associated costs: systems require maintenance and computational power, and auditors must be employed to monitor them and investigate when a potential fraud case is identified [12]. The expense of a false positive, in misclassifying a legitimate transaction as fraud, is typically far less than that of a false negative [14]. Insufficient study has been performed on the disproportionate nature of these costs, with attention typically focussing on the traditional classification performance methods outlined in Section 2.1. Considering the accuracy of each fraud detection method, focus should be on achieving an optimum balance for each technique such that the expense is smallest. Research specifically focused on finding this balance would add significant real-world value to financial fraud detection.

Given the diversity of common categories of fraud it would be useful to have some form of generic framework that could apply to more than one fraud category. Such a framework could be used to study the differences between various types of fraud, or even specific details such as differentiating between stolen and counterfeit credit cards [3]. A ubiquitous model could also be used to determine which specific fraud detection method is applicable given the problem domain. This approach has been investigated slightly with response surface methodology [30], but more detailed research is desirable.

## 4 Conclusion

Fraud detection is an important part of the modern finance industry. In this research, we have investigated the current practices in financial fraud detection using intelligent approaches, both statistical and computational. Though their performance differed, each technique was shown to be reasonably capable at detecting various forms of financial fraud. In particular, the ability of CI methods such as neural networks and support vector machines to learn and adapt to new situations is highly effective at defeating the evolving tactics of fraudsters.

There are still many aspects of intelligent fraud detection that have not yet been the subject of research. Some types of fraud, as well as some data mining methods, have been superficially explored but require future study to be completely understood. There is also the opportunity to examine the performance of existing methods by using customisation or tuning, as well as the potential to study cost benefit analysis of computational fraud detection. Finally, further research into the differences between each type of financial fraud could lead to a generic framework which would greatly enhance the scope of intelligent detection methods for this problem domain.

**References**

1. Bai B, Yen J, and Yang X (2008) False financial statements: characteristics of China's listed companies and CART detecting approach. *International Journal of Information Technology & Decision Making* **7**, 339-59.
2. Bermúdez L, Pérez J, Ayuso M, Gómez E, and Vázquez F (2008) A Bayesian dichotomous model with asymmetric link for fraud in insurance. *Insurance: Mathematics and Economics* **42**, 779-86.
3. Bhattacharyya S, Jha S, Tharakunnel K, and Westland JC (2011) Data mining for credit card fraud: A comparative study. *Decision Support Systems* **50**, 602-13.
4. Bose I and Wang J (2007) Data mining for detection of financial statement fraud in Chinese Companies. Paper presented at the International Conference on Electronic Commerce, Administration, Society and Education, Hong Kong, 15-17 August 2007.
5. Cecchini M, Aytug H, Koehler GJ, and Pathak P (2010) Making words work: Using financial text as a predictor of financial events. *Decision Support Systems* **50**, 164-75.
6. Duman E and Ozcelik MH (2011) Detecting credit card fraud by genetic algorithm and scatter search. *Expert Systems with Applications* **38**, 13057-63.
7. Glancy FH and Yadav SB (2011) A computational model for financial reporting fraud detection. *Decision Support Systems* **50**, 595-601.
8. Holton C (2009) Identifying disgruntled employee systems fraud risk through text mining: A simple solution for a multi-billion dollar problem. *Decision Support Systems* **46**, 853-64.
9. Hoogs B, Kiehl T, Lacomb C, and Senturk D (2007) A genetic algorithm approach to detecting temporal patterns indicative of financial statement fraud. *Intelligent Systems in Accounting, Finance and Management* **15**, 41-56.
10. Humpherys SL, Moffitt KC, Burns MB, Burgoon JK, and Felix WF (2011) Identification of fraudulent financial statements using linguistic credibility analysis. *Decision Support Systems* **50**, 585-94.
11. Jans M, van der Werf JM, Lybaert N, and Vanhoof K (2011) A business process mining application for internal transaction fraud mitigation. *Expert Systems with Applications* **38**, 13351-9.
12. Kirkos E, Spathis C, and Manolopoulos Y (2007) Data mining techniques for the detection of fraudulent financial statements. *Expert Systems with Applications* **32**, 995-1003.
13. Koh HC and Low CK (2004) Going concern prediction using data mining techniques. *Managerial Auditing Journal* **19**, 462-76.
14. Ngai E, Hu Y, Wong Y, Chen Y, and Sun X (2011) The application of data mining techniques in financial fraud detection: A classification framework and an academic review of literature. *Decision Support Systems* **50**, 559-69.

15. Paasch CA (2010) In *Credit card fraud detection using artificial neural networks tuned by genetic algorithms*. Vol. pp. HONG KONG UNIV. OF SCI. AND TECH.(HONG KONG),
16. Panigrahi S, Kundu A, Sural S, and Majumdar AK (2009) Credit card fraud detection: A fusion approach using Dempster–Shafer theory and Bayesian learning. *Information Fusion* **10**, 354-63.
17. Pinquet J, Ayuso M, and Guillen M (2007) Selection bias and auditing policies for insurance claims. *Journal of Risk and Insurance* **74**, 425-40.
18. Quah JT and Sriganesh M (2008) Real-time credit card fraud detection using computational intelligence. *Expert Systems with Applications* **35**, 1721-32.
19. Ravisankar P, Ravi V, Raghava Rao G, and Bose I (2011) Detection of financial statement fraud and feature selection using data mining techniques. *Decision Support Systems* **50**, 491-500.
20. Sánchez D, Vila M, Cerda L, and Serrano J-M (2009) Association rules applied to credit card fraud detection. *Expert Systems with Applications* **36**, 3630-40.
21. Sohl JE and Venkatachalam A (1995) A neural network approach to forecasting model selection. *Information & Management* **29**, 297-303.
22. Vatsa V, Sural S, and Majumdar AK (2005) A game-theoretic approach to credit card fraud detection. In *Information Systems Security*. Vol. pp. 263-76. Springer.
23. Viaene S, Ayuso M, Guillen M, Van Gheel D, and Dedene G (2007) Strategies for detecting fraudulent claims in the automobile insurance industry. *European Journal of Operational Research* **176**, 565-83.
24. Whitrow C, Hand DJ, Juszczak P, Weston D, and Adams NM (2009) Transaction aggregation as a strategy for credit card fraud detection. *Data Mining and Knowledge Discovery* **18**, 30-55.
25. Wu SX and Banzhaf W (2008) Combatting financial fraud: a coevoutionary anomaly detection approach. In *Proceedings of the 10th annual conference on Genetic and evolutionary computation*. (ed.), Vol. pp. 1673-80, ACM.
26. Yang W-S and Hwang S-Y (2006) A process-mining framework for the detection of healthcare fraud and abuse. *Expert Systems with Applications* **31**, 56-68.
27. Yeh I and Lien C-h (2009) The comparisons of data mining techniques for the predictive accuracy of probability of default of credit card clients. *Expert Systems with Applications* **36**, 2473-80.
28. Yue D, Wu X, Wang Y, Li Y, and Chu C-H (2007) A review of data mining-based financial fraud detection research. In *Wireless Communications, Networking and Mobile Computing, 2007. WiCom 2007. International Conference on*. (ed.), Vol. pp. 5519-22, IEEE Press.
29. Zhang D and Zhou L (2004) Discovering golden nuggets: data mining in financial application. *Systems, Man, and Cybernetics, Part C: Applications and Reviews, IEEE Transactions on* **34**, 513-22.
30. Zhou W and Kapoor G (2011) Detecting evolutionary financial statement fraud. *Decision Support Systems* **50**, 570-5.